# Upper limit on efficiency of electron deflection in bent crystals

V.M. Biryukov

Institute for High Energy Physics, 142281 Protvino, Russia

**Abstract**

Basing on experiments at SLAC, MAMI, CERN, and on Monte Carlo simulations, we assess the physical limits on efficiency and reachable angles of electron deflection in bent crystals in the energy range from sub-GeV to sub-TeV. We find that applications of electron channeling for beam deflection are possible, limited by maximal deflection angle of 2-3 mrad at sub-GeV, 1-2 mrad at multi-GeV, and 0.5-1 mrad at sub-TeV energy. This opens opportunities for crystal-assisted electron beam extraction from accelerator rings.

## 1. Introduction

Channeling of positive particles in bent crystals has been studied quite substantially with protons, heavy ions, positrons, and even short-lived particles over the recent four decades [1-20] since the idea of Tsyganov [21] to apply crystal channeling for beam bending. Several applications of bent crystals have been realized for the steering of proton beams at accelerators. At IHEP Protvino, crystal extraction systems work since the 1980s delivering beam to particle physics experiments [22-26]. Channeling efficiency of nearly 100% is obtained at intensities in the order of $10^{12}$ proton/s [25]. More applications are proposed for many positive particles (from positrons to heavy ions, from protons to $\Lambda_c$ baryons and tau leptons) in a wide energy range from sub-GeV to multi-TeV [27-35].

The technique strongly developed in the studies at JINR, CERN, FNAL, and IHEP, can lead to interesting applications at the LHC, such as crystal collimation making the collider cleaner by an order of magnitude [30,31], or crystal extraction [32,33] opening the way to variety of fixed-target experiments [28,29,34,35].

More recent development is the studies with negative particles channeled in bent crystals. Several experiments were made on negative beam deflection by means of bent crystal channeling at the SPS, REFER (Hiroshima), SLAC, and MAMI (Mainz) with electrons and negative pions with energy from 150 MeV to 150 GeV [36-39]. These experiments have shown that it is possible to deflect negative particles by a fraction of milliradian with efficiency up to order of 10%.

We would like to summarize the data of several experiments and Monte Carlo simulations with negative particles for a broad energy range and various curvatures of crystals. It is important to understand what parameters of crystals are optimal for steering of negative particles, what deflection angles are reachable in possible applications, and what efficiencies are really feasible. This would allow us to estimate the perspectives of crystal applications for steering of negative particles at accelerators.

Channeling of positive projectiles in bent crystals is reasonably well understood theoretically. Analytical models and Monte Carlo simulations are able to predict the results of experiments [11]. To be trapped in channeling mode, the incident particle must be within the Lindhard critical angle $\theta_L$. Respectively, from the incident beam with divergence $\Phi$ only a fraction $\theta_L/\Phi$ fits the Lindhard angle. The particles that fit the critical angle are also restricted by so-called surface acceptance $A_S$ which takes into account that atomic planes are actually atomic layers of certain thickness.

The bending of a crystal causes a loss of some channeled particles represented by a bending dechanneling factor $A_B$. When the crystal bending radius becomes as small as the Tsyganov critical radius $R_C$, channeling disappears, $A_B=0$. Finally, the particles trapped in bent crystal undergo scattering on the crystal constituents, nuclei and electrons. This scattering leads to a gradual loss, dechanneling, over the crystal length $L$, represented by the exponential factor with characteristic dechanneling length $L_D$. The overall efficiency $F$ of particle deflection by a bent crystal can be estimated as

$$F \approx (\theta_L/\Phi) \cdot A_S \cdot A_B \cdot \exp(-L/L_D). \qquad (1)$$

The Lindhard critical angle is about the same for positive and negative particles channeled in crystal planes. The Tsyganov critical radius $R_C$ in bent crystal planes is also about the same for particles of both sign. The radical difference between positive and negative projectiles in crystal planes is their dechanneling length $L_D$.

Unlike in the case of positive projectiles which, once trapped, are moving in a rarified electronic gas between the positively-charged atomic planes, the negative projectiles are attracted to the atomic planes and have to experience strong nuclear scattering which makes their dechanneling length $L_D$ (i.e., occupation length in the channeled states) pretty short.

## 2. Dechanneling angle $\theta_D$ for negative particles in bent crystals

The most important data obtained in the experiments was the dechanneling lengths $L_D$ measured for negative particles in bent crystals at different energies in various crystals for the range of bending radii. The dechanneling length for negative particles is about two orders of magnitude shorter than this length for positive particles at the same energy in the same crystal. Applications of electron channeling are limited by strong scattering of particles on the crystal nuclei which leads to rapid dechanneling of the trapped particles.

The coefficients at the exponent in Eq. (1) determine how much of the particles incident at crystal can be trapped in channeling mode. They depend on the divergence of particles and on the curvature of crystal. In a circular accelerator, channeling efficiency would also depend on the multiplicity factor for particles circulating in the accelerator ring and having multiple chances to enter the crystal [40,41]. In a very favorable situation when you have either a very low divergence of the particles incident at crystal, or a great multiplicity of particle encounters with crystal in the ring, or both, the overall coefficient at the exponent can be close to 100%.

However, even if you trap all 100% of the particles initially, you inevitably lose many of them when you try to deflect the trapped particles, because of the dechanneling process described by the exponent. For large deflection angles, the factor limiting the efficiency is this exponent $\exp(-L/L_D)$. By analogy with dechanneling length which limits the length that particle can traverse in a channeled state in crystal, we introduce a characteristic dechanneling angle $\theta_D = L_D/R$ which limits the deflection angle $\theta$ that particle can be bent with:

$$\exp(-L/L_D) = \exp(-R\theta/L_D) = \exp(-\theta/\theta_D). \quad (2)$$

Thus, the exponent $\exp(-\theta/\theta_D)$ sets the fundamental upper limit on the efficiency of particle deflection at sufficiently large angle $\theta$. Like the dechanneling length in bent crystals, the introduced dechanneling angle $\theta_D$ may be a function of crystal curvature and beam energy.

By using beam of low divergence, $\Phi \leq \theta_L$, and crystal with low curvature, $A_B \approx 1$, you can maximize the overall coefficient at the exponent $\exp(-L/L_D)$. However, the dechanneling losses are unavoidable. Optimization of electron deflection efficiency for large angles means first of all the optimization of the dechanneling angle which is a function of crystal curvature and beam energy.

|  |  | $E$ (GeV) | $R$ (cm) | $E/R$ (GeV/cm) | $L_D$ (μm) | $L_D/R$ (μrad) |
|---|---|---|---|---|---|---|
| SLAC | 2016 | 3.35 | 15 | 0.22 | 55.4 | 369 |
| SLAC | 2016 | 4.2 | 15 | 0.28 | 45.2 | 301 |
| SLAC | 2016 | 6.3 | 15 | 0.42 | 65.3 | 435 |
| SLAC | 2016 | 10.5 | 15 | 0.7 | 57.5 | 383 |
| SLAC | 2016 | 14 | 15 | 0.93 | 55.8 | 372 |
| Biryukov | 2007 | 50 | 41.67 | 1.2 | 110 | 264 |
| MAMI | 2014 | 0.855 | 3.35 | 0.26 | 19.2 | 573 |
| Mazzolari et al. | 2014 | 0.855 | 0.8 | 1.07 | 6.9 | 864 |
| Mazzolari et al. | 2014 | 0.855 | 1.6 | 0.53 | 13.6 | 850 |
| Mazzolari et al. | 2014 | 0.855 | 3.3 | 0.26 | 19.5 | 591 |
| Mazzolari et al. | 2014 | 0.855 | 6.7 | 0.13 | 25.5 | 381 |
| Mazzolari et al. | 2014 | 0.855 | 13.4 | 0.06 | 35.5 | 265 |
| SPS | 2013 | 150 | 1920 | 0.08 | 930 | 48 |

**Table 1** Dechanneling length measured in crystals bent with radius $R$ for negative particles at SLAC, MAMI, SPS [37-39], and simulated by Monte Carlo codes [38,42]. We calculate in this table the dechanneling angle $\theta_D = L_D/R$ for every case.

Table 1 summarizes the data from different experiments [37-39] and simulations [42-44] for the dechanneling length of electrons in bent crystals for different energies and bending radii. We calculate in this table the dechanneling angle $\theta_D = L_D/R$ for every case.

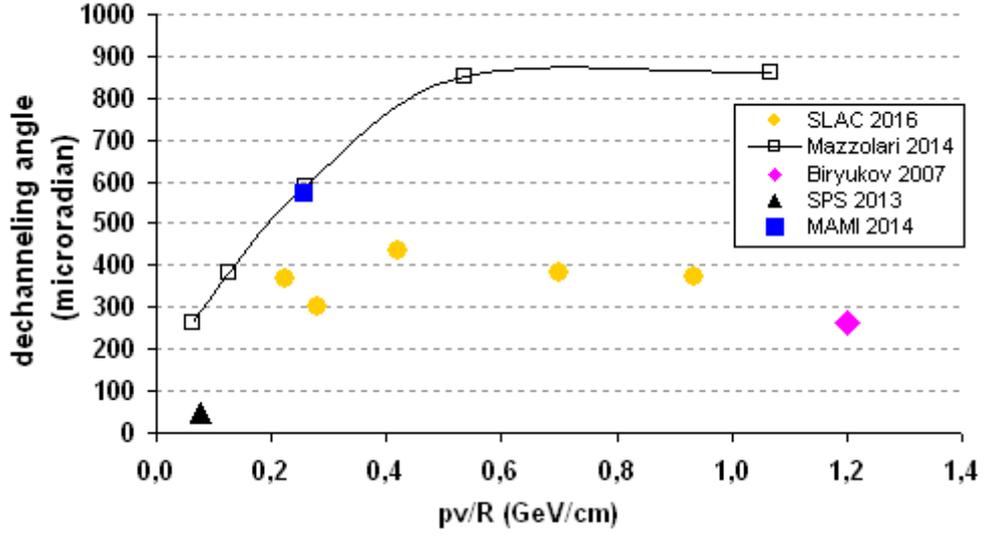

**Fig. 1** Dechanneling angle $L_D/R$ measured for negative particles in bent crystals at SLAC, MAMI, CERN and found in Monte Carlo simulations, plotted as a function of the crystal curvature $pv/R$.

Fig. 1 shows the deflection angle $L_D/R$ measured for negative particles in bent crystals at SLAC, MAMI, CERN [37-39] and found in Monte Carlo simulations [42-44], plotted as a function of the crystal curvature $pv/R \approx E/R$. Strictly speaking, we need to use $pv$ (particle momentum times velocity) instead of $E$ (energy) in the formulas for channeling and scattering. However, in this paper we always consider ultrarelativistic cases, so we often replace $pv \approx E$. We see that the dechanneling angle $\theta_D$ is stable over a broad interval of crystal curvature. According to SLAC results, the optimal range of crystal curvature $pv/R$ is as wide as from ~0.2 to ~1 GeV/cm. The average dechanneling angle $\theta_D$ at SLAC energy of 3-14 GeV calculated for the studied range of crystal curvature equals 370±50 µrad.

Fig. 2 shows the deflection angle $\theta_D = L_D/R$ measured for negative particles in bent crystals at SLAC, MAMI, CERN and found in Monte Carlo simulations, plotted as a function of the particle energy. We see that at the lower energy, below 1 GeV, the dechanneling angle $\theta_D$ is larger, about 600 µrad. At much higher energy the dechanneling angle $\theta_D$ appears much lower, down to ~100 µrad in this plot. On one hand, this high-energy result is influenced by the low curvature of the crystals used at CERN. However, the fundamental reason for lower dechanneling angle of electrons at greater energy is the deviation of electron dechanneling length from linear dependence on energy.

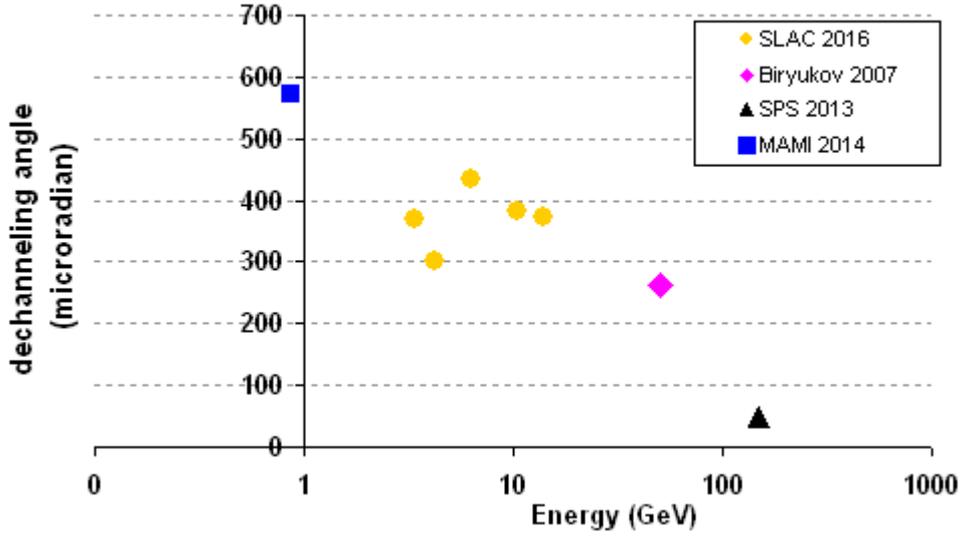

**Figure 2** Dechanneling angle $L_D/R$ measured for negative particles in bent crystals at SLAC, MAMI, CERN and found in Monte Carlo simulations, plotted as a function of the particle energy.

Table 2 shows the dechanneling length measured in straight crystals for negative particles at the energies from 0.855 MeV to 150 GeV at several accelerator centers [37-39,45] in 1984-2013, together with prediction of Monte Carlo code CATCH [42-44].

|         |      | $E$ (GeV) | $L_D/E$ (µm/GeV) | $L_D/25R_C$ (µrad) |
|---------|------|-----------|------------------|--------------------|
| MAMI    | 2014 | 0.855     | 22.6             | 542                |
| KhFTI   | 1984 | 1.2       | 23.3             | 560                |
| SLAC    | 2016 | 3.35 to 14| 15.4             | 370                |
| Biryukov| 2007 | 50        | 6.6              | 158                |
| SPS     | 2013 | 150       | 6.2              | 149                |

**Table 2** Dechanneling length measured in straight crystals for negative particles at the energies from 0.855 MeV to 150 GeV at several accelerator centers in 1984-2013, together with Monte Carlo prediction.

The Tsyganov critical radius $R_C$ grows linearly with energy. If the dechanneling length of electron would grow linearly with energy as well, then the dechanneling angle $\theta_D$ would be a constant independent of energy. From Table 2 it is clear that the electron dechanneling length deviates from linear dependence on energy. The ratio $L_D/E$ is not constant; it reduces by factor ~4 as the electron energy grows from ~1 GeV to ~100 GeV. Respectively, dechanneling angle reduces with energy. Therefore, the deflection angle achievable for negative particles at 100 GeV is factor ~4 lower than it is at 1 GeV. We attribute this deviation from linear dependence to logarithmic factor known also for positive particles [11].

We estimate the characteristic dechanneling angle $\theta_D$ as ~600 µrad at sub-GeV, ~400 µrad at multi-GeV, and only 150 µrad at sub-TeV energy. Comparing Tables 1 and 2, we notice that the highest value for the experimentally measured dechanneling angle $\theta_D$ can be approximately estimated as

$$\theta_D \approx L_D/25R_C, \qquad (3)$$

where $L_D$ is for straight crystals and $R_C$ is the Tsyganov critical radius. We provide $L_D/25R_C$ in Table 2.

## 3. Efficiency limits for electron deflection

We show in Fig. 3 the predicted upper limit $\exp(-\theta/\theta_D)$ on deflection efficiency for negative particles in bent crystals, plotted as a function of the deflection angle for sub-GeV, multi-GeV (3-14 GeV), and sub-TeV (~100 GeV) energies.

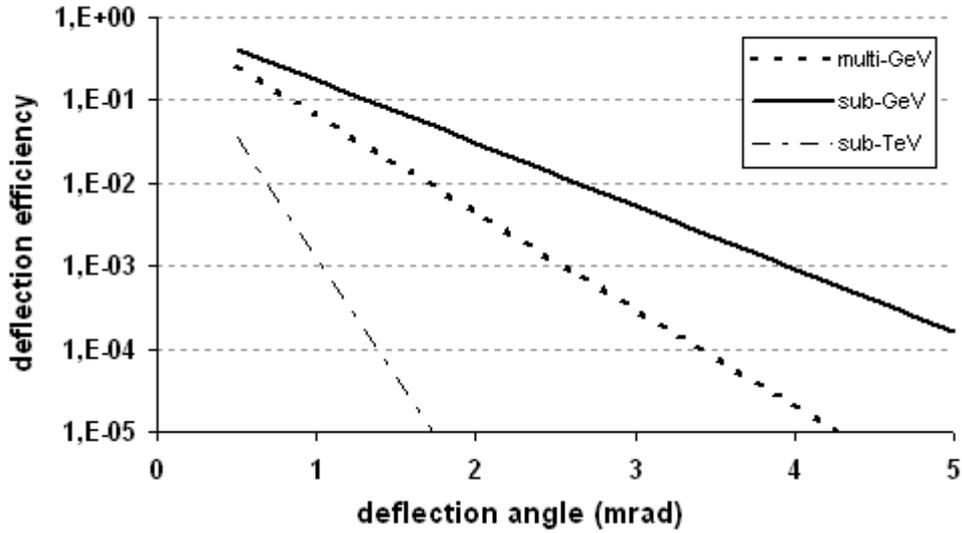

**Figure 3** The predicted upper limit on deflection efficiency for negative particles in bent crystals, plotted as a function of the deflection angle. Curves for sub-GeV, 3-14 GeV, and ~100 GeV energies (top to bottom).

It might be possible to reach the electron deflection angle of 1 mrad with efficiency 10-20% at sub-GeV energy, and of 5-10% at multi-GeV energy. The deflection of 2 mrad would be possible only with efficiency of the order of 3% at sub-GeV energy and just 0.5% or lower at multi-GeV energy. At the energy above 100 GeV the reachable angle for deflection of negative particles is just 0.5 mrad (with efficiency order of ~4%) to 1 mrad (with efficiency of just ~0.1%).

## 4. Practical records for proton deflection and conclusions for electrons

One can estimate the maximum deflection angle that can be practically achieved for negative particles in a different way. For protons, the largest angle of deflection that was used in practical applications at accelerators amounted to 150 mrad [46]. At IHEP, a 150-mrad bent crystal was used to create a test area. This "crystal beamline" worked in parallel with other experimental set-ups consuming no power and delivering 70-GeV protons with intensity $10^6$ p/s to particle experiments. Crystal efficiency was as low as ~0.01%.

Besides this 150-mrad bent crystal of 100-mm length operated since 1994, the other examples in the IHEP experience were 85-mrad bent crystal [22] and 89-mrad bent crystal of 85 mm length [47] used for extraction onto PROZA and BEC, respectively, physics experiments in the period of 1989–1999. For the LHC purpose, that practical record of proton deflection at 150 mrad (i.e. 9º, equivalent to about 32 T-m!) has stimulated the proposal [48] to bend the LHC protons and ions by a huge angle from 1º to 20º in the energy range of 0.45–7 TeV by means of a bent crystal for calibration of CMS or ATLAS calorimeters *in situ* by the LHC beam of precisely known energy.

For negative particles the dechanneling length is a factor of ≈ 50 shorter that it is for protons. This can be seen, for instance, from the measurements of $L_D$ in Si(111) at similar energy of ~1 GeV for electrons at MAMI (Mainz) [38] and KhFTI (Kharkov) [45] where $L_D ≈$ 20 μm/GeV was measured, and for protons at PNPI (Gatchina) where $L_D ≈$ 1 mm/GeV was measured [49,11].

Ref. [50] proposed a simple explanation for this factor 50. Nuclear scattering (responsible for dechanneling of negative particles) is much stronger than electronic one (responsible for dechanneling of positive particles). A nucleus of charge $Ze$ scatters a particle an angle $\Delta\theta$ factor of $Z$ larger than a single charge $e$ does. That is, the angle diffusion $(\Delta\theta)^2/\Delta z$ is a factor of $Z^2$ faster per unit length. Taking into account that, per each atom, only $n_e$ valence electrons contribute to electronic scattering of protons channeled in crystal planes, ref.[50] estimated that nuclear scattering provides a dechanneling rate a factor of $Z^2/n_e$ larger. Of course, this is a simplified model which neglects other differences in scattering and dechanneling of negative and positive particles but emphasizes the major factor.

For Silicon, $Z^2/n_e ≈ 50$ because in semiconductors the number of valence electrons $n_e ≈$ 4. Respectively, the largest practical angle of electron deflection would be a factor $Z^2/n_e ≈$ 50 lower than the 150-mrad practical record for protons:

(150 mrad) · $n_e/Z^2$ ≈ (150 mrad)/50 = 3 mrad             (4)

Remember that the efficiency of electron bending 3 mrad would be about as low as the efficiency of proton bending 150 mrad, i.e. only order of 0.01% for multi-GeV beam.

## 5. Conclusion

We conclude that factor $\exp(-\theta/\theta_D)$ is the limit on the efficiency of electron deflection at angle $\theta$. Here, $\theta_D = L_D/R$ is the parameter well known from the experiments and simulations. This limit on the efficiency gives a hope to design interesting applications on deflection of electrons with bent crystals, such as beam extraction from accelerator rings, for instance from DAFNE ring at LNF [27].